\documentclass{svjour3}                     
\usepackage{graphicx}
\journalname{Gen. Relat. Grav.}
\begin{document}

\title{Boson star and dark matter}
\titlerunning{Boson star and dark matter}

\author{R. Sharma, S. Karmakar and S. Mukherjee}
\authorrunning{Sharma, Karmakar, Mukherjee}
\institute{R. Sharma \at St. Joseph$'$s College, Darjeeling-734 104, India\\
\email{ranjan$\_$sh@hotmail.com}
\and
S. Karmakar\at Department of Physics, North Bengal University, Darjeeling-734 013,\\
\email{karma@iucaa.ernet.in}
\and
S. Mukherjee \at
Inter-University Centre for Astronomy and Astrophysics, Post bag
4, Pune- 411 007, India
 and
Astrophysics and Cosmology Research Unit,
 School of Mathematical Science, University of KwaZulu-Natal,
Private Bag X54001, Durban 4000, South Africa\\
\email{sailom@iucaa.ernet.in}
}

\date{Received: date / Accepted: date}

\maketitle

\begin{abstract}
Bound states of complex scalar fields (boson stars) have long been
proposed as possible candidates for the dark matter in the
universe. Considerable work has already been done to study various
aspects of boson stars. In the present work, assuming a particular
anisotropic matter distribution, we solve the
Einstein-Klein-Gordon equations with a cosmological constant to
obtain bosonic configurations by treating the problem
geometrically. The results are then applied to problems covering a
wide range of masses and radii of the boson stars and the relevant
self interaction parameters are calculated. We compare our results
with earlier treatments to show the applicability of the
geometrical approach.
 \keywords{Exact solution \and Einstein-Klein-Gordon equation \and Boson star \and Dark matter.}
\end{abstract}

\section{Introduction}
\label{intro}
 According to recent cosmological observations, approximately 96\% of the
total material content of the universe is exotic in nature out of
which 73\% is gravitationally repulsive (dark energy) and the
remaining 23\% is attractive in nature but exists in the form of
dark matter\cite{Matos02,Garnavich,Perlmutter}. The quest for such
exotic matter in the universe remains an outstanding problem in
cosmology and astrophysics. In some models, massive axion of mass
$m\sim 10^{-5}~$eV is considered as a standard dark matter
particle (with no self-interaction). These massive axions may
collapse to form compact objects known as axion stars of masses
comparable to the mass of a planet. The survival of such axion
stars has been questioned by Seidel and Suen\cite{Seidel}.
Moreover, these small objects cannot explain the observed
abundance of dwarf galaxies and the dark matter density profiles
in the core of galaxies. The view that dark matter in the galactic
halos could be due to the presence of scalar field confogurations
has been considered by many\cite{Guzman,Matos01}, (see also
Ref.\cite{Schunck} for a recent review). Hu {\em et al}\cite{Hu}
showed that the large scale structure of the galactic halo could
be explained by dark matter composed of ultralight scalar
particles of mass $m\sim 10^{-22}~$eV. Lee and Koh\cite{koh} have
investigated boson stars with a self-interacting scalar field as
the model for galactic halos. These results are consistent with
the observation that the mass of a scalar particle should be
greater than $10^{-28}~$eV not to disturb radial
stability\cite{Gleiser}.

The large abundance of dark energy in the universe is also a field
of active research. It appears that the dark energy could be well
described by a cosmological constant\cite{Liddle}. In this paper
we will accept this interpretation. The effect of the cosmological
constant on the mass or radius of a compact boson star will be
negligible. However, we will work in the general framework where
the cosmological constant will be retained\cite{Bohmer}.

Historically, the concept of a gravitationally bound state of
scalar particles, was proposed by Kaup\cite{Kaup} and Ruffini and
Bonazzola\cite{Ruffini}. Since a boson star is prevented from
gravitational collapse by Heisenburg uncertainty principle, the
maximum mass of a boson star with no self-interaction, is very
small. However, Colpi {\em et al}\cite{Colpi} showed that a
self-interaction of the scalar particles could yield compact
objects of masses comparable to neutron stars. In the absence of
direct observational evidence of fundamental scalar particles
constituting a boson star, various models have been proposed to
study the gross features of such hypothetical
objects\cite{Jetzer,Jetzer1,Henriques,Henriques1,Hafizi,Capozziello,Ho}.
One of the main motivations to study these objects is to
understand its possible role in explaining the dark matter in the
universe, although boson stars made up of Higg's particles,
dilatons and other scalar particles may also exist.The choice of
the potential corresponding to the self-interaction is very
important in the calculations of such models. A quartic type
potential ($\Lambda\Phi^4$) is, in general, used for the
self-interaction\cite{Colpi,Capozziello}, though some works on
higher order self-interactions\cite{Ho} and different forms such
as $\cosh$-type potential\cite{Matos02,Alcubierre} have also been
reported in the literature. The motivation for choosing a certain
potential form is to obtain configurations which are consistent
with astrophysical and cosmological observations. For example,
assuming a $\cosh$-type potential of the form $V(\Phi)
=V_{0}[\cosh(\lambda\sqrt{8\pi G}\Phi) -1]$, Alcubierre {\em et
al} \cite{Alcubierre} showed that one can construct a
self-gravitating scalar object or oscillaton of critical mass
$M_{crit} \sim 10^{12}~M_{\odot}$ which is comparable with the
dark matter content of a galactic halo where, the parameters
have values $\lambda\sim 20.28$, $V_{0}\sim (3\times
10^{-27}~m_{pl})^4$, $m_{\Phi}\sim1.1\times10^{-23}$~eV. Henriques
and Mendes\cite{Henriques} considered a combined boson-fermion
star and showed that it is possible to generate a wide variety of
configurations ranging from objects of atomic sizes and masses of
the order of $10^{-15}~M_{\odot}$ to objects having galactic
masses of the order of $10^{13}~M_{\odot}$ and radii extending
upto a few light years. Of particular interest is a configuration
where the fermion core and the bosonic halo could form a
supermassive compact object of mass $M \sim 10^{13}~M_{\odot}$ for
the bosonic particle mass $m_{b} \sim 10^{-32}$~GeV and fermionic particle mass $m_{f}
\sim 10^{-5}$~GeV. The method that one usually follows in such
calculations is that one makes a choice for the potential and then
solves the Einstein-Klein-Gordon system numerically by fixing the
values of the coupling constant and the value of the scalar field
at the origin. It will be interesting if we assume the mass and
radius of a boson star first and then look for the potential
corresponding to the assumed configuration. In the present work,
we explore this alternative approach, where the mass of the scalar
particle and the coupling constant get determined by the field
equations. Essentially we will solve the Einstein-Klein-Gordon
system with an energy momentum tensor corresponding to a
self-interacting complex scalar field in the presence of a
cosmological constant. Current observational studies done by
High-Z Supernova Team\cite{Garnavich} and Supernova Cosmological
Project group\cite{Perlmutter} clearly indicate that, though
small, we nevertheless need a positive cosmological constant to
explain the current acceleration of the universe expansion. The
current value of the cosmological constant is estimated to be
$\Lambda_c \sim 3\times 10^{-56}$~cm$^{-2}$\cite{Padmanabhan}. In
our model, it helps us to construct a star with a proper boundary.
In general, in the standard calculation of a boson star, one
imposes the condition that the  scalar field vanishes
asymptotically which makes it difficult to define the radius
properly (usually defined as where $90$\% of the mass is
contained). However, a cosmological constant term makes it
possible to define the radius in a conventional way (as in a
fermion star), i.e., where the radial pressure vanishes. Similar
observations may also be found in some other
papers\cite{Perlmutter,AGReiss}. In a recent paper\cite{Schunck1}
the radius of the dark halo is described as the distance where
pressure is equal to the cosmological constant $\Lambda_{c}$ of
the order of magnitude $10^{-29}~$gm/cm$^3$ or alternatively, the
cosmic microwave background radiation (CMBR) which is about
$10^{-34}~$gm/cm$^3$ and the corresponding radii are $1~$Mpc for
$\Lambda_{c}$ and $100~$Mpc for the CMBR. The exterior of the halo
is described by the Schwarzschild de-Sitter spacetime. Note that
boson stars have already been studied in de-Sitter as well as
anti-de-Sitter universe\cite{Astefanesei}.

Our paper is organized as follows. The boson star model is
developed in Section 2 where, by making use of an ansatz for one
of the metric potentials for an anisotropic matter distribution,
we generate a solution for the Einstein-Klein-Gordon system. In
Section 3, we utilize this solution to obtain different boson star
configurations and determine the nature of the self interaction.
We discuss our results in Section 4 and outline the physical
implications of our studies along with some comments on the
earlier work on the possibility of the formation of a boson
star\cite{Seidel02} and on its stability\cite{Seidel03} under
radial perturbations.

\section{Boson star configuration}
\label{sec:1} We write the energy-momentum tensor for a scalar
field in the form
\begin{equation}
T_{ij} = \frac{1}{2}\left(\Phi_{i}^{*}\Phi_{j}+
\Phi_{i}\Phi_{j}^{*}\right) -\frac{1}{2} g_{ij}\left[g^{mn}
\Phi_{m}^{*} \Phi_{n} + V(|\Phi|^2)\right]\label{Eq1},
\end{equation}
where $\Phi(r,t)$ is a complex scalar field and $V(|\Phi|^2)$ is
the potential of self-interaction of the scalar field. The scalar
field is assumed to be of the form
\begin{equation}
\Phi(r,t) = \phi(r)e^{-i\omega t}\label{Eq2},
\end{equation}
which guarantees a spherically symmetric static matter
distribution. We write the metric for the spherically symmetric
space-time in the standard coordinates
\begin{equation}
ds^{2} = -e^{2\gamma (r)}dt^{2}+e^{2\mu (r)}dr^{2}+r^{2}(d\theta
^{2}+\sin ^{2}\theta d\phi ^{2})\label{Eq3},
\end{equation}
where $\gamma(r)$ and $\mu(r)$ are to be determined. The
Einstein's field equations (with $c=1$) are then obtained as
\begin{eqnarray}
8\pi G\rho \equiv \left[\frac{\left(1-e^{-2\mu}\right)}{r^2}+\frac{2\mu'e^{-2\mu}}{r}-\Lambda_{c}\right]
=\frac{1}{2}\left[\omega^2 e^{-2\gamma}\phi^2+{\phi'}^2 e^{-2\mu}+V(|\Phi|^2)\right]\label{Eq4},\\
8\pi G p_{r} \equiv \left[\frac{2\gamma'e^{-2\mu}}{r}-\frac{\left(1-e^{-2\mu}\right)}{r^2}+\Lambda_{c}\right]
=\frac{1}{2}\left[\omega^2 e^{-2\gamma}\phi^2+{\phi'}^2 e^{-2\mu}-V(|\Phi|^2)\right]\label{Eq5},\\
8\pi G p_{\perp}\equiv \left[ e^{-2\mu}\left(\gamma''+{\gamma'}^2-\gamma'\mu'+\frac{\gamma'}{r}-
\frac{\mu'}{r}\right)+\Lambda_{c}\right] =\frac{1}{2} \left[\omega^2 e^{-2\gamma}\phi^2-{\phi'}^2 e^{-2\mu}-V(|\Phi|^2)\right]\label{Eq6},
\end{eqnarray}
where $\Lambda_c$ is a cosmological constant term. A prime ($'$) here
denotes differentiation with respect to the radial coordinate
$r$. The Klein-Gordon equation
\begin{equation}
\left[\fbox{} - \frac{dV}{d|\Phi|^2}\right]\Phi = 0 \label{Eq7},
\end{equation}
for the line element (\ref{Eq3}) takes the form
\begin{equation}
\phi''+\left(\frac{2}{r}+\gamma'-\mu'\right)\phi'+e^{2(\mu-\gamma)}
\omega^2\phi = e^{2\mu} \frac{dV}{d\phi^2}\phi\label{Eq8}.
\end{equation}
Eq.~(\ref{Eq4})-(\ref{Eq6}) together with Eq.~(\ref{Eq8}) comprise
the system of equations describing the boson star.

Note that by analogy with an anisotropic fluid, we may identify
Eq.~(\ref{Eq4})-(\ref{Eq6}) as density($\rho$), radial
pressure($p_{r}$) and tangential pressure($p_{\perp}$) equations,
respectively\cite{Gleiser}. It then follows that in the interior
of a boson star, pressure is anisotropic due to the term
${\phi'}^2 e^{-2\mu}$. Clearly $p_{r}
> p_{\perp}$ in this model. If $\Delta = 8 \pi G(p_{r}- p_{\perp})
$
 is the measure of pressure anisotropy, i.e.,
\begin{equation}
 \Delta = {\phi'}^2 e^{-2\mu}\label{Eq9},
\end{equation}
then Eq.~(\ref{Eq5}) and Eq.~(\ref{Eq6}) may
be combined to yield
\begin{equation}
\gamma ^{\prime \prime }+{\gamma ^{\prime }}^{2}-\gamma ^{\prime}\mu ^{\prime }-\frac{\gamma ^{\prime }}{r}
-\frac{\mu ^{\prime}}{r}-\frac{\left( 1-e^{2\mu }\right) }{{r}^2}+\Delta e^{2\mu } = 0\label{Eq10},
\end{equation}
which is a non linear second order differential equation with two
coupled functions $\gamma(r)$ and $\mu(r)$. To solve Eq.~(\ref{Eq10}) we
first make use of an ansatz for one of the metric potentials given by
Vaidya and Tikekar \cite{VT01} in the form
\begin{equation}
e^{2\mu } =
\frac{1+kr^{2}/R^{2}}{1-r^{2}/R^{2}}\label{Eq11},
\end{equation}
where $R$ and $k$ are dimensionless parameters.
Eq.~(\ref{Eq10}) then gets the form
\begin{equation}
(1+k-k x^2)\Psi_{xx} + k x \Psi_{x} + k (k + 1)\Psi + \frac{\Delta
R^2 (1+k - k x^2)^2}{(1-x^2)} \Psi = 0\label{Eq12},
\end{equation}
where we introduced the following transformations
$$ \Psi = e^{\gamma},~~~ x^{2} = 1-\frac{r^2}{R^2}.$$
To solve Eq.~(\ref{Eq12}) the radial dependence of the anisotropic
parameter $\Delta$ needs to be specified. We make a choice for the
anisotropic parameter\cite{sk}
\begin{equation}
\Delta = \frac{\alpha k^2(1-x^2)}{R^2(1+k-k x^2)^2}\label{Eq13},
\end{equation}
where $\alpha > 0$ is a constant. Eq.~(\ref{Eq12}) then gets the form
\begin{equation}
(1-z^2)\Psi_{zz} + z\Psi_{z} + [k(1+\alpha) + 1]\Psi  = 0\label{Eq14},
\end{equation}
where $z = \sqrt{k/(k+1)} x$. Using the properties of Gegenbauer function and Tschebyscheff polynomial, a general solution of Eq.~(\ref{Eq14}) may be obtained as\cite{SNB01}
\begin{equation}
\Psi(z) = e^{\gamma } = A\bigg[{\frac{\cos [(\beta+1)\zeta +\delta
]}{\beta+1}}-{\frac{ \cos [(\beta-1)\zeta
+\delta]}{\beta-1}}\bigg]\label{Eq15},
\end{equation}
where, $\beta = \sqrt{k(1+\alpha) + 2}$, $\zeta =\cos ^{-1}z$, and $A$ and
$\delta$ are constants. The energy density ($\rho$), radial pressure
($p_r$), tangential pressure ($p_{\perp}$) and the anisotropic
parameter ($\Delta$) are then obtained as
\begin{eqnarray}
\rho &=& \frac{1}{8 \pi G R^2 (1-z^2)} \bigg[ 1 + {2 \over (k + 1)(1 - z^2)} \bigg]-\frac{\Lambda_{c}}{8\pi G}\label{Eq16}, \\
p_{r} &=& - \frac{1}{8 \pi G R^2 (1-z^2)} \bigg[ 1 + {2z \Psi_{z}\over(k + 1) \Psi} \bigg]+ \frac{\Lambda_{c}}{8\pi G}\label{Eq17}, \\
p_{\perp} &=& p_{r} - \Delta\label{Eq18}, \\
\Delta &=& \frac{\alpha k}{8 \pi G R^2} \bigg[ {(k + 1)(1 - z^2) - 1 \over
(k + 1)^2(1 - z^2)^2} \bigg]\label{Eq19}.
\end{eqnarray}
Thus all the physical parameters can be obtained once the geometry
of the star is specified.

Combining equations.~(\ref{Eq9}), (\ref{Eq11}) and (\ref{Eq19})
and integrating we also obtain a solution for the scalar wave
function in the form
\begin{equation}
\sigma = \sigma_{0} + \sqrt{\frac{\alpha k}{2}}\sin^{-1}z,
\label{Eq20}
\end{equation}
where $\sigma_{0}$ is an integration constant.

\subsection{Determination of the Potential}
To analyze the form of the potential, we now express Eq.~(\ref{Eq4})-(\ref{Eq6}) and Eq.~(\ref{Eq8}) in dimensionless forms as
\begin{equation}
\bar{\rho} \equiv \frac{\left(1-e^{-2\mu}\right)}{r_{*}^2}+\frac{2\mu'e^{-2\mu}}{r_{*}}-\Lambda_{*}
= \left[\Omega^2 e^{-2\gamma}\sigma^2+{\sigma'}^2 e^{-2\mu}+\frac{4\pi G}{m^2}V(\sigma)\right]\label{Eq21},
\end{equation}
\begin{equation}
\bar{p_{r}}\equiv \frac{2\gamma'e^{-2\mu}}{r_{*}}-\frac{\left(1-e^{-2\mu}\right)}{r_{*}^2}+\Lambda_{*}
=\left[\Omega^2 e^{-2\gamma}\sigma^2+{\sigma'}^2 e^{-2\mu}-\frac{4\pi G}{m^2}V(\sigma)\right]\label{Eq22},
\end{equation}
\begin{equation}
\bar{p_{\perp}} \equiv e^{-2\mu}\left(\gamma''+{\gamma'}^2-\gamma'\mu'
+ \frac{\gamma'}{r_{*}}-\frac{\mu'}{r_{*}}\right)+\Lambda_{*} = \left[\Omega^2
e^{-2\gamma}\sigma^2-{\sigma'}^2 e^{-2\mu}-\frac{4\pi
G}{m^2}V(\sigma)\right]\label{Eq23},
\end{equation}
\begin{equation}
\sigma'' +(\gamma' - \mu' +\frac{2}{r_{*}})\sigma'+ \Omega^2
e^{2(\mu -\gamma)}\sigma = \frac{1}{2}e^{2\mu}(\frac{4\pi
G}{m^2})\frac{dV}{d\sigma}\label{Eq24},
\end{equation}
where we have made the following rescaling: $r_{*} = r m$, $R_{*}=R m$, $\sigma = \sqrt{4\pi G}\phi$,
 $\Omega =\omega/m$, $\Lambda_{*}=\Lambda_{c}/m^2$, $\bar{\rho}=8\pi G \rho$, $\bar{p_{r}}= 8 \pi G p_{r}$
  and  $\bar{p_{\perp}}= 8 \pi G p_{\perp}$. Here $m$ is the mass of the scalar particle and a prime
  now denotes differentiation with respect to $r_{*}$.

We rewrite the Eq.~(\ref{Eq24}) in the form
\begin{eqnarray}
\frac{(1-nz^2)}{n^2z^2R_{*}^2}\sigma_{zz} +
\left[-\frac{1}{n^2R_{*}^2z^3}-\frac{2}{nzR_{*}^2}+
\frac{(1-nz^2)}{n^2z^2R_{*}^2}\gamma_{z}-\frac{(1-nz^2)}{n^2z^2R_{*}^2}\mu_{z}\right]\sigma_{z}
\nonumber \\ +\Omega^2e^{2(\mu-\gamma)}\sigma = \frac{1}{2}
e^{2\mu}\frac{4\pi G}{m^2} \frac{dV}{d\sigma}, \label{Eq25}
\end{eqnarray}
where $n=(k+1)/k$. Defining the potential in dimensionless form as
\begin{equation}
\tilde{V} = 4\pi G R^2 V = \frac{4\pi R_{*}^2}{m^2 m_{pl}^2}
V,\label{Eq26}
\end{equation}
we write Eq.~(\ref{Eq25}) in the form
\begin{eqnarray}
\frac{(1-nz^2)}{n^2z^2}\sigma_{zz} +
\left[-\frac{1}{n^2z^3}-\frac{2}{n
z}+\frac{(1-nz^2)}{n^2z^2}\gamma_{z}-
\frac{(1-nz^2)}{n^2z^2}\mu_{z}\right]\sigma_{z} \nonumber  \\
+{\tilde{\Omega}}^2e^{2(\mu-\gamma)}\sigma = \frac{1}{2}
e^{2\mu}\frac{d\tilde{V}}{d\sigma}, \label{Eq27}
\end{eqnarray}
where, $\tilde{\Omega}=\Omega R_{*}$. The physical quantities are then obtained as
\begin{eqnarray}
\tilde{\rho} = \tilde{\Omega}^2 e^{-2\gamma}\sigma^2 +\frac{(1-n z^2)}{n^2 z^2}e^{-2\mu}\sigma_{z}^2+\tilde{V}, \label{Eq28}\\
\tilde{p_{r}} = \tilde{\Omega}^2 e^{-2\gamma}\sigma^2 +\frac{(1-n z^2)}{n^2z^2}e^{-2\mu}\sigma_{z}^2-\tilde{V},\label{Eq29}\\
\tilde{p_{\perp}} = \tilde{\Omega}^2e^{-2\gamma}\sigma^2 -\frac{(1-n z^2)}{n^2z^2}e^{-2\mu}\sigma_{z}^2-\tilde{V},\label{Eq30}\\
\tilde{\Delta} = 2\frac{(1-nz^2)}{n^2z^2}e^{-2\mu}\sigma_{z}^2. \label{Eq31}
\end{eqnarray}

Combining the two sets of equations, i.e., Eq.~(\ref{Eq16}) - (\ref{Eq19}) and Eq.~(\ref{Eq28}) - (\ref{Eq31}), we get
\begin{eqnarray}
\tilde{\rho}\equiv
\frac{1}{(1-z^2)}\left[1+\frac{2}{(1+k)(1-z^2)}\right] -\tilde{\Lambda_{c}}=
\tilde{\Omega}^2 e^{-2\gamma}\sigma^2 +\frac{(1-n z^2)}{n^2 z^2}e^{-2\mu}\sigma_{z}^2+\tilde{V}, \label{Eq32}\\
\tilde{p_{r}}\equiv
-\frac{1}{(1-z^2)}\left[1+\frac{2z\gamma_{z}}{(1+k)}\right]  +\tilde{\Lambda_{c}}=
\tilde{\Omega}^2 e^{-2\gamma}\sigma^2 +\frac{(1-n z^2)}{n^2
z^2}e^{-2\mu}\sigma_{z}^2-\tilde{V},\label{Eq33}\\
\tilde{p_{\perp}}\equiv \tilde{p_{r}} - \tilde{\Delta}  =
\tilde{\Omega}^2
e^{-2\gamma}\sigma^2 -\frac{(1-n z^2)}{n^2z^2}e^{-2\mu}\sigma_{z}^2-\tilde{V},\label{Eq34}\\
\tilde{\Delta}\equiv {\alpha k} \bigg[ {(k + 1)(1 - z^2) - 1 \over (k +
1)^2(1 - z^2)^2} \bigg]=
2\frac{(1-nz^2)}{n^2z^2}e^{-2\mu}\sigma_{z}^2,  \label{Eq35}
\end{eqnarray}
where $\tilde{\Lambda_{c}}=8\pi G R^{2}\Lambda_{c}$. Eq.~(\ref{Eq32})-(\ref{Eq35}) thus help us to obtain our physical quantities from two different perspectives - one from the grometry and the other from the energy-momentum part of the field equations.

\subsection{Boundary conditions}
To calculate the physical quantities of the geometrical part of the field equations, we impose the following boundary conditions.\\
(1) At the boundary of the star $(r = b)$ the interior solution is
matched to the external Schwarzschild-de Sitter metric which gives
\begin{eqnarray}
e^{2\gamma(r=b)} &=&  \left(1 - \frac{2M}{b}-\frac{\Lambda_{c} b^2}{3}\right), \label{Eq36}\\
e^{2\mu(r=b)} &=& \left(1 - \frac{2M}{b}-\frac{\Lambda_{c}
b^2}{3}\right)^{-1} \label{Eq37}.
\end{eqnarray}
We note that the effect of $\Lambda_{c}$ for highly compact
objects like neutron stars is negligible in this model. However,
for large objects whose compactness is very low, the contribution
due to $\Lambda_{c}$ can not be ignored.

(2) The radial pressure $\tilde{p_{r}}$ should
vanish at the boundary which gives
\begin{equation}
\frac{\Psi_{z}(z_{b})}{\Psi(z_{b})} = -\frac{(1+k)}{2z_{b}}\left[1-\tilde{\Lambda_{c}}(1-z_{b}^2)\right],
\label{Eq38}
\end{equation}
where $z_{b}^2 = (k/(k +1))(1 - b^2/R^2)$.

For given values of mass and radius, Eq.~(\ref{Eq36})-(\ref{Eq38})
may be used to calculate the values of the constants $A$, $R$ and
$\delta$ for assumed values of the curvature parameter $k$ and the
anisotropic parameter $\alpha$. Thus all the physical quantities
can be obtained once the geometry is specified.

\section{Numerical results}
Using the results of Section 2, we now choose different bosonic configurations and calculate the interaction parameters. The method is the following.

\begin{figure}
\includegraphics[width=0.75\textwidth]{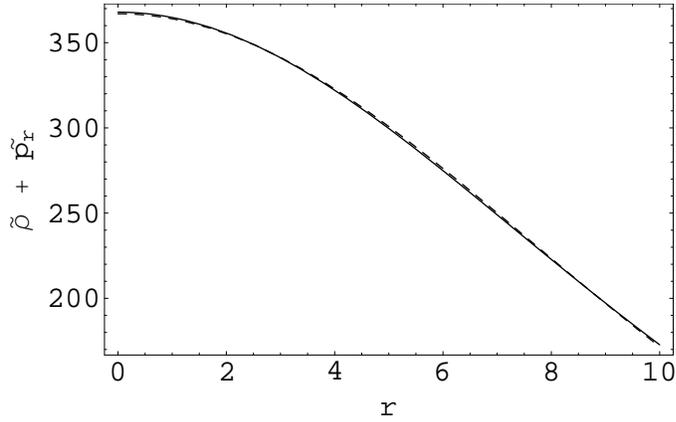}
\caption{($\tilde{\rho} +\tilde{p_{r}}$) plotted against radius $r~$(km). The solid curve originates from the geometry and the
dotted curve is derived from the matter part of the field
equations. We took, $M = 1~M_{\odot}$, $b=10~$km and $\alpha=1$.}
\label{fig1}
\end{figure}

For a given mass $M$ and radius $b$, we calculate the constants from the boundary conditions, viz., $R$ (from Eq.~(\ref{Eq37})), $\delta$ (from Eq.~(\ref{Eq38})) and $A$ (from Eq.~(\ref{Eq36})). We combine Eq.~(\ref{Eq32}) and (\ref{Eq33}) to obtain
\begin{eqnarray}
\tilde{\rho} +\tilde{p_{r}}\equiv \frac{1}{(1-z^2)}\left[1+\frac{2}{(1+k)(1-z^2)}\right]
-\frac{1}{(1-z^2)}\left[1+\frac{2z\gamma_{z}}{(1+k)}\right] = \nonumber \\
2\tilde{\Omega}^2 e^{-2\gamma}\sigma^2 +\frac{2(1-nz^2)}{n^2
z^2}e^{-2\mu}\sigma_{z}^2,\label{Eq39}
\end{eqnarray}
where there are two unknown parameters, $\tilde{\Omega}$ and $\sigma_{0}$. We plot ($\tilde{\rho} +\tilde{p_{r}}$) against the radial coordinate $r$ from the centre to the boundary of the star using the geomerty part (left hand side) of Eq.~(\ref{Eq39}). By
suitably choosing the values of $\tilde{\Omega}$ and $\sigma_{0}$, we fit this curve with the one obtained from the left hand side of Eq.~(\ref{Eq39}). This has been shown in Figure \ref{fig1}.

\begin{figure}
\includegraphics[width=0.75\textwidth]{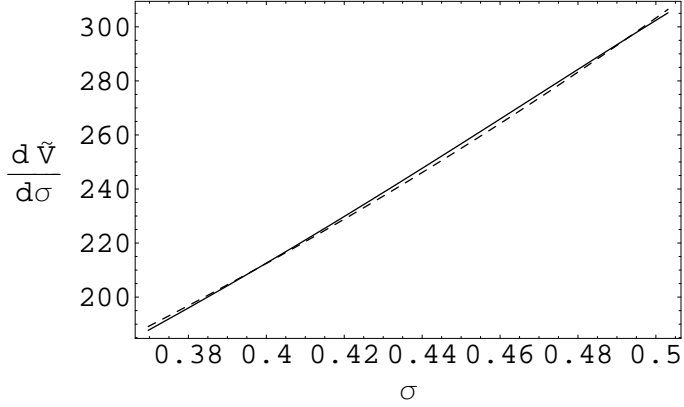}
\caption{$\frac{d\tilde{V}}{d\sigma}$ plotted against $\sigma$.
The solid curve originates from the geometry and the
dotted curve is derived from the matter part of the field
equations. We took, $M = 1~M_{\odot}$, $b=10~$km and $\alpha=1$.}
\label{fig2}
\end{figure}

We now choose the scalar wave potential in the form
$$V=V_{0}+m^2|\Phi|^2+\lambda|\Phi|^4,$$
where the constant potential term $V_{0}$, the coupling constant $\lambda$ and the mass of the scalar particle $m$ are yet to be fixed. To this end, we rewrite the potential $V$ in dimensionless form as
\begin{equation}
\tilde{V} = 4\pi G R^2 V = \tilde{V_{0}}+ R_{*}^2\sigma^2+\Lambda R_{*}^2 \sigma^4,
\label{Eq40}
\end{equation}
where $\Lambda= \lambda/{4\pi G m^2}$ and $\tilde{V_{0}}=  4\pi G R^2 V_{0}$.  Differentiating Eq.~(\ref{Eq40}) we get
\begin{equation}
\frac{d\tilde{V}}{d\sigma} = 2R_{*}^2\sigma + 4\Lambda R_{*}^2
\sigma^3. \label{Eq41}
\end{equation}
We now plot $\frac{d\tilde{V}}{d\sigma}$ against $\sigma$ using Eq.~(\ref{Eq27}) and Eq.~(\ref{Eq41}) as shown in Figure \ref{fig2}. This fixes the values of $R_{*}$ and $\Lambda$. The mass of the scalar particle is then obtained from the relation $m = R_{*}/R$.

\begin{figure}
\includegraphics[width=0.75\textwidth]{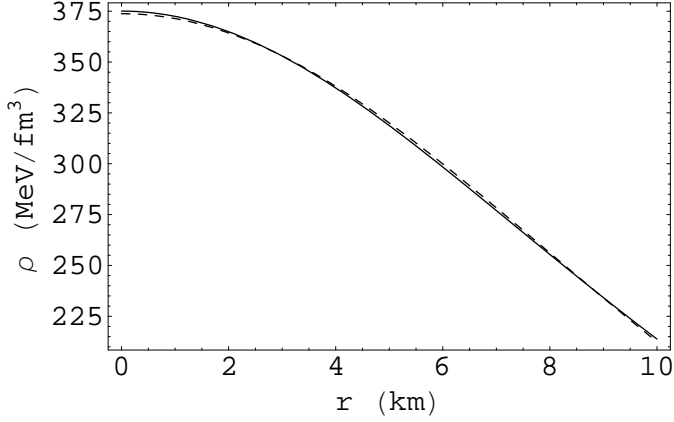}
\caption{Energy density $\rho$ plotted against radius $r$. The solid curve originates from the geometry and the
dotted curve is derived from the matter part of the field
equations. We took, $M = 1~M_{\odot}$, $b=10~$km and $\alpha=1$.}
\label{fig3}
\end{figure}

\begin{figure}
\includegraphics[width=0.75\textwidth]{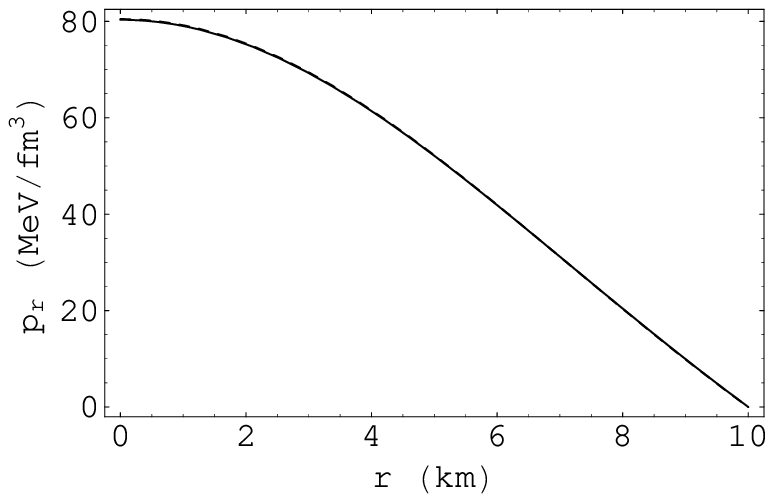}
\caption{Radial pressure $p_{r}$ plotted against radius $r$. The solid curve originates from the geometry and the
dotted curve is derived from the matter part of the field
equations. We took, $M = 1~M_{\odot}$, $b=10~$km and $\alpha=1$.}
\label{fig4}
\end{figure}

\begin{figure}
\includegraphics[width=0.75\textwidth]{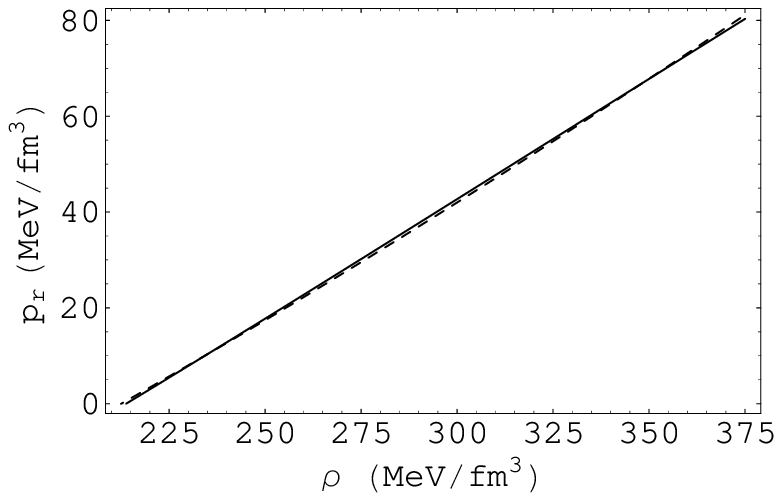}
\caption{Radial pressure $p_{r}$ plotted against density $\rho$. The solid curve originates from the geometry and the
dotted curve is derived from the matter part of the field
equations. We took, $M = 1~M_{\odot}$, $b=10~$km and $\alpha=1$.}
\label{fig5}
\end{figure}

Finally making a suitable choice for the constant $\tilde{V_{0}}$, the energy density
$\tilde{\rho}$ and the radial pressure $\tilde{p_r}$ are plotted against the
radius $r$ as shown in Figure \ref{fig3} and Figure
\ref{fig4}, respectively. The variation of radial pressure with
density (EOS) both from the geometric part and the matter part are
shown in Figure \ref{fig5}.

Thus we get a complete description of the boson star. Following the technique, we have considered a wide variety of bosonic configurations and the results have been compiled in Table \ref{bstab1}.
\begin{table}
\begin{center}
\begin{tabular}{|l|c|c|c|c|c|c|c|c|r|}
\hline 
$\alpha$ & $\delta$ & $A$ & $-\sigma_{0}$ & $\tilde{\Omega}$ &$R_{*}$&  $\Lambda$ & $\tilde V_{0}$& $V_{0}$(Mev/fm$^{3}$) & $m$(eV)\\ \hline
\multicolumn{10}{|c|}{{\bf Case I:}  $k=100$, $M=1M_{\odot}$, $b=15~ km$ and $R=305~km$}\\ \hline
0.5 & 2.738 & 40.094 & 6.993 & 29.710 & 24.463 & 1.599 & 42.2 & 27.4 & $1.6\times10^ {-11}$\\ \hline
1.0& 2.644 & 48.478 & 9.965 & 24.948 &18.168 & 0.648 & 60.8 & 39.5 & $1.2\times10^ {-11}$\\ \hline
2.0 & 2.409 &62.550 & 14.229 & 23.179 & 9.710 & 1.572 & 90.7 & 58.9 & $6.3\times10^ {-12}$\\ \hline
\hline
\multicolumn{10}{|c|}{{\bf Case II:} $k=100$, $M=1M_{\odot}$, $b=10~ km$ and $R=155.68~km$}\\ \hline
0.2 & 2.647 & 31.133 & 4.462 & 33.320 & 28.546 & 6.352 & 27.8 & 69.3 & $3.6\times10^ {-11}$\\ \hline
0.5 & 2.615 & 35.962 & 6.943 & 23.651 &19.670 & 2.083 & 38.1 & 94.8  &  $2.5\times10^ {-11}$\\ \hline
1.0 &2.511 & 43.231 & 9.899 & 19.733 & 14.090 & 1.055 & 54.8 & 136.5 &  $1.8\times10^ {-11}$\\ \hline
1.5 & 2.386 & 49.661 & 12.196 & 18.536 & 9.786 & 1.392 & 69.6 & 173.5 &  $1.2\times10^ {-11}$\\ \hline
2.0 & 2.257 & 55.446 & 14.149 & 18.234 & 3.638 & 11.305 & 81.6 & 203.4 &  $4.6\times10^ {-12}$\\ \hline
\hline
\multicolumn{10}{|c|}{{\bf Case III:} $k=100$, $M=1.5M_{\odot}$, $b=10~ km$ and $R=113.25~km$}\\ \hline
0.5 & 2.378 & 29.789 & 6.822 & 14.787 & 10.679 & 4.348 & 34.6 & 162.9 & $1.8\times10^ {-11}$\\ \hline
1.0 & 2.252 & 35.337 & 9.738 & 12.079 &  5.874  & 4.083 & 49.1 & 231.7 & $1.0\times10^ {-11}$\\ \hline
1.2 & 2.195 & 37.384 & 10.700 & 11.592 & 3.185 & 10.789 & 54.8 & 258.0 & $5.6\times10^ {-12}$\\ \hline
\hline
\multicolumn{10}{|c|}{{\bf Case IV:} $k=2$, $M=1M_{\odot}$, $b=10~ km$ and $R=28.58~km$}\\ \hline
0.5 & 2.301 & 1.099 & 0.207 & 3.590 & 2.006 & 7.12 &1.66 & 122.73 & $1.4\times10^ {-11}$\\\hline
1.0 & 2.274 & 1.243 & 0.376 & 2.928 & 1.740 & 2.49 & 1.94 & 143.43 & $1.2\times 10^{-11}$\\\hline
2.0 & 2.157 & 1.511 & 0.693 & 2.621 & 1.089 & 2.17 & 2.58 & 190.75& $7.5\times10^{-12}$\\\hline
2.2 & 2.128 & 1.561 &0.753 & 2.604 & 0.907 & 2.90 & 2.70& 199.62 & $6.3\times10^{-12}$ \\ \hline
2.5 & 2.085 & 1.635 & 0.840 & 2.590 & 0.529 & 8.11 & 2.87 & 212.19 & $3.6\times10^{-12}$ \\ \hline
\hline
\end{tabular}
\caption{Results of different bosonic configurations.}
\label{bstab1}
\end{center}
\end{table}

\section{Discussion}
\begin{figure}
\includegraphics[width=0.75\textwidth]{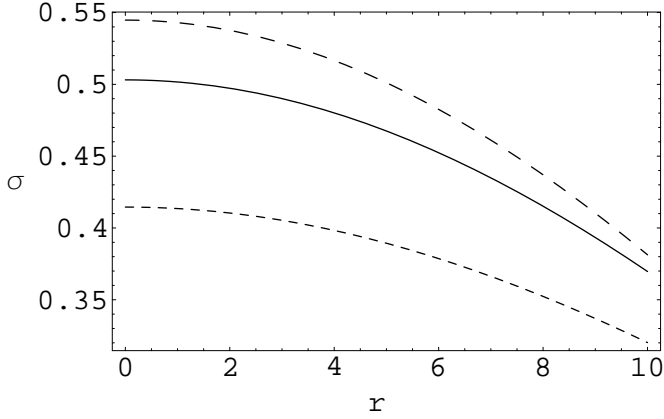}
\caption{Scalar field $\sigma$ plotted against radius $r~(km)$. Dotted,
solid and dashed curves are for $\alpha$ equal to $0.5$, $1$ and $1.5$
respectively.}
\label{fig6}
\end{figure}

\begin{figure}
\includegraphics[width=0.75\textwidth]{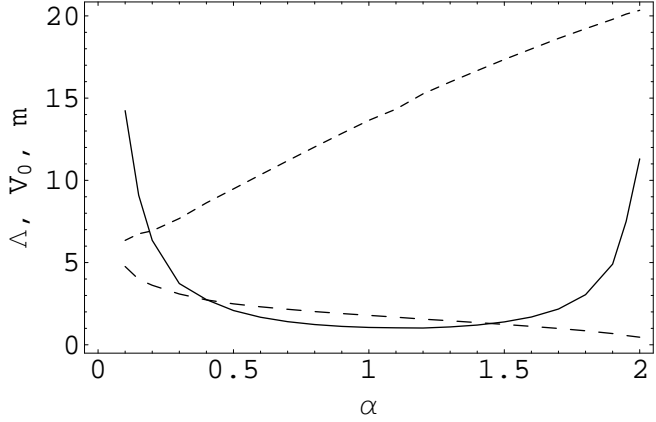}
\caption{Variations of $\Lambda$ (solid line), $V_{0}/10$ in $Mev/fm^3$
(dotted line) and $m/10^{-11}$ in $eV$ (long dashed line) with
$\alpha$.} \label{fig7}
\end{figure}

\begin{figure}
\includegraphics[width=0.75\textwidth]{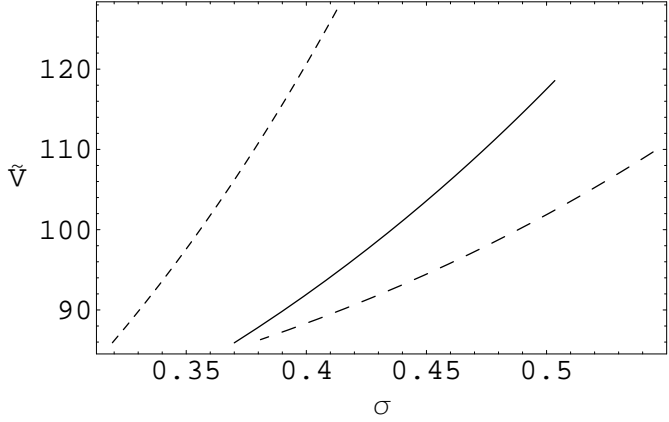}
\caption{Scalar field potential $\tilde{V}$ plotted against scalar field $\sigma$. Dotted,
solid and dashed curves are for $\alpha$ equal to $0.5$, $1$ and $1.5$
respectively.}\protect\label{fig8}
\end{figure}

\begin{figure}
\includegraphics[width=0.75\textwidth]{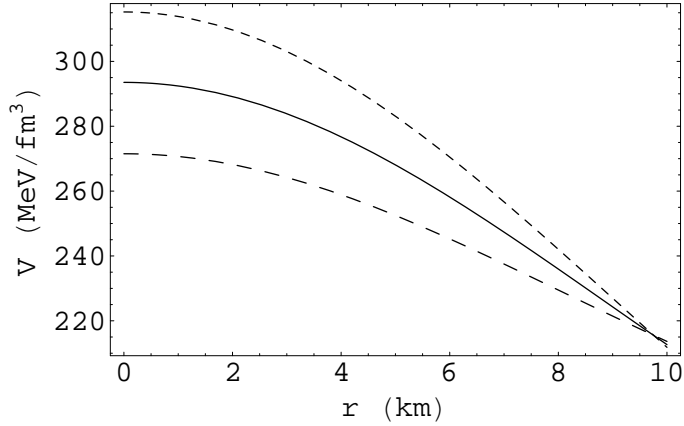}
\caption{Scalar field potential $\tilde{V}$ plotted against radius
$r$. Dotted, solid and dashed curves are for $\alpha$ equal to
$0.5$, $1$ and $1.5$ respectively.}\protect\label{fig9} 
\end{figure}

The geometric approach developed in this paper helps us to study the form of the potential of a wide variety of bosonic configurations. The key features of our studies are given below:
\begin{itemize}
\item Considering a boson star of mass $M = 1~M_{\odot}$ and radius $b = 10$ km, we have shown the variation of scalar
field $\sigma$ against radius $r$ for three different values of
the anisotropic parameter $\alpha$ as shown in Figure \ref{fig6}.
The scalar field decreases smoothly towards the boundary from the
centre and is nodeless, as desired. Possible radial excitations of
the scalar field configurations have not been considered here.

\item For given values of $M$, $b$ and $k$, the mass of the scalar particle $m$ decreases and the constant potential term
$V_{0}$ increases with the increase of the anisotropic factor $\alpha$. The coupling constant $\Lambda$ first decreases,
remains almost constant for a specific range of $\alpha$, and then increases. These variations are shown in Figure \ref{fig7}.

\item Note that $\alpha$ cannot be made arbitrarily large. In our
model, the maximum value of $\alpha$ decreases with increasing
compactness for a given $k$. However, for the same compactness it
increases if $k$ is made small.

\item In this model we have considered a quartic type potential only, however, higher order terms in the potential may also be
 considered. Variations of the potential $\tilde{V}$ against the scalar field $\sigma$ and with respect to $r$
 inside the stellar interior have been shown in Figure \ref{fig8} and Figure \ref{fig9}, respectively.

\item Our method can be used to study a wide variety of bosonic configurations. In Table \ref{bstab2} we have compiled
some of our results which are comparable with the results
obtained by earlier workers.

\item {\bf Boson star of mass of the order of a planet:} The model can
be used to describe a boson star of mass of the order of a planet. For
example, if we consider a star of mass $M=10^{-6}~M_{\odot}$ and
radius $b=10^{-5}$~km, the mass of the scalar particle turns out to be $m\sim 10 ^{-5}~$eV. In CDM model, it is observed
 that massive axions of similar order of masses can collapse to form compact object of mass $M\sim {0.6m_{p}}^2/m \sim 10^{-6}~M_{\odot}$ \cite{Seidel}.

\item {\bf Axion star:} Stable axion stars have masses less than
$0.846~M_{\odot}$ and radii greater than $20.5~$km with mass of the
scalar field $m\sim 10 ^{-10}~$eV, close to the lower bound of
axions\cite{MSchunck}. For a star of mass $0.8~M_{\odot}$ and radius
$20.5~$km, we get the mass of the scalar particle in the same order.

\item {\bf Boson star in galactic centre:}  Some galactic centres are supposed to contain super-massive compact dark
objects. It is believed that these ultra-compact objects are
either supermassive black holes or very compact clusters of
 stellar size black holes\cite{Genzel}. Observational data seem to allow the possibility that these could well be
 bosonic stars. In a boson star, the radius $b \geq M({m_{Pl}^{-2})}$\cite{Torres}. Data collected for the central object
 of a galaxy by Genzel {\em et al}\cite{Genzel} show that for a boson star model as in Ref.\cite{Torres}, the mass and
 radius of the central object should be $M\sim 10^6 M_{\odot}$ and $b \sim 10^7~$km, respectively. The mass of the
 corresponding scalar particle turns out to be $m\sim 10^{-17}~$eV. These results are consistent with our calculations.

\item {\bf Boson star of galactic scale:} In this model, we find that for a scalar particles of mass $m\sim 10 ^{-23}~$eV
 and constant potential $V_{0}\sim 10^{-22}$~Mev/fm$^3$, a boson star of mass $M\sim 10^{12}~M_{\odot}$ and radius
 $b\sim 10^{13}~$km may be obtained. These values are comparable to the scalar field dark matter model of
 Ref.~\cite{Alcubierre}, where $m_{\Phi}\sim1.1\times10^{-23}~$eV and $V_{0}\sim 10^{-25}~$Mev/fm$^3$ for a potential of
  the form $V(\Phi) =V_{0}[\cosh(\lambda\sqrt{8\pi G}\Phi) -1]$. The size of the galaxy of mass $\sim 10^{12}~M_{\odot}$ was about $10^{13}~$km. Thus the results are in
good agrement with the present model.

\item We have shown that a wide variety of boson stars may be obtained in this model by either scaling or by considering different compactness. Obviously not all of these stars will be realistic. One also needs to consider the stability of such configurations under radial perturbations. This may provide constraints on the mass of the scalar
field\cite{koh}, and therefore, the mass of the boson star. For radial
stability, the mass of boson should be $m \geq 10^{-28}~$eV\cite{koh,Gleiser}. In this model, with such scalar particle masses, one can describe a boson star of mass and radius as high as $M \sim 10^{18}~M_{\odot}$ and $b \sim 10^{19}$~km, respectively.
\end{itemize}

\begin{table}
\begin{center}
\begin{tabular}{|l|c|c|c|r|}
\hline
 $M$ ($M_{\odot}$) & $b$(km)  & $V_{0}$ (Mev/fm$^3$) & $\Lambda$ & $m$ (eV)  \\ \hline
 $10^{-6}$ & $10^{-5}$ & $9.619\times 10^{13}$ & 2.08 & $2.46 \times 10^{-5}$ \\
$0.8$ & $20.5$ & $8.4$ & 1.34 & $ 10^{-11}$\\
$10^{6}$ & $10^{7}$   & $9.619\times 10^{-11}$ & 2.08 & $2.46 \times 10^{-17}$ \\
$10^{12}$ & $10^{13}$ & $9.619\times 10^{-23}$ & 2.08 &  $2.46 \times 10^{-23}$ \\
$10^{18}$ & $10^{19}$ & $9.619\times 10^{-35}$ & 2.08 & $2.46 \times 10^{-29}$ \\
\hline\hline
\end{tabular}
\caption{Variations of scalar particle masses for different boson star configurations with $\alpha = 0.5$ and $k = 100$.}
\label{bstab2}
\end{center}
\end{table}

To conclude, although the present model is based on a particular
geometry, depending on two parameters (Vaidya-Tikekar model), and
a simple anisotropic distribution, the resulting configurations,
at least a class of them, may be physically relevant. This is
confirmed by comparing the results of our geometrical approach
with earlier results, obtained by direct numerical integration.
Whether boson stars can actually account for the dark matter, or
at least a part of it, will require extensive analysis of the
observational results of the CMB radiation, the Supernova data and
other relevant astronomical results. The stability of the boson
star configurations under radial perturbations also remains an
interesting issue. For scalar fields without self-interaction,
extensive numerical calculations done by Seidel and Suen
\cite{Seidel02} have shown interesting stability properties of the
configurations for finite perturbations. Whether boson stars with
self-interacting fields have similar behaviour remains to be
checked. It may be also be pointed out that Seidel and
Suen\cite{Seidel03} have also suggested a mechanism, the
gravitational cooling, which permits the scalar field to get rid
of the excess kinetic energy so that a bound configuration can be
formed (instead of a virialised cloud). These results make the
possibility of the formation of boson stars look more likely. We
hope to take up these issues elsewhere.

\begin{acknowledgements}
SK would like to thank the IUCAA, Pune, India and also the IUCAA
Reference Centre, North Bengal University, India for providing
facilities during the course of this work. SM is thankful to the
Astrophysics and Cosmology Research Unit of the University of
Kwazulu- Natal and the National Institute for Theoretical Physics,
South Africa at Durban, where a part of the work was done and to
S. D. Maharaj and S. Roy for discussions.
\end{acknowledgements}


\begin{thebibliography}{}
\bibitem{Matos02}T. Matos and L. A. Ure$\acute{n}$a-L$\ddot{o}$pez, {\em Int. J. Mod. Phys. D} {\bf13}, 2287 (2004).
\bibitem{Garnavich} P. M. Garnavich {\em et al}, {\em Astrophys. J.} {\bf509}, 74 (1998).
\bibitem{Perlmutter} S. Perlmutter {\em et al}, {\em Nature} {\bf391}, 51 (1998).
\bibitem{Seidel} E. Seidel and W. Suen, {\em gr-qc/9412062}.
\bibitem{Guzman} F. S. Guzm$\acute{a}$n and T. Matos, {\em Class. Quantum Grav.} {\bf17}, L9 (2000).
\bibitem{Matos01} T. Matos and L. A. Ure$\acute{n}$a-L$\ddot{o}$pez, {\em Phys. Rev. D} {\bf63}, 63506 (2001).
\bibitem{Schunck} F. E. Schunck and E. W. Mielke, {\em Class. Quantum Grav.} {\bf 20}, R301 (2003).
\bibitem{Hu} W. Hu, R. Barkana and A. Gruzinov, {\em Phys. Rev. Lett.} {\bf85}, 1158 (2000).
\bibitem{koh} J. Lee and I. Koh, {\em Phys. Rev. D} {\bf53}, 2236 (1996).
\bibitem{Gleiser} M. Gleiser, {\em Phys. Rev. D} {\bf38}, 2376 (1988).
\bibitem{Liddle} A. R. Liddle, {\em Mon. Not. R. Astron. Soc.} {\bf290}, 533 (1997).
\bibitem{Bohmer} C. G. B$\ddot{o}$hmer, {\em Gen. Relat. Grav.} {\bf36}, 1039 (2004).
\bibitem{Kaup} D. J. Kaup, {\em Phys. Rev.} {\bf172}, 1331  (1968).
\bibitem{Ruffini} R. Ruffini and S. Bonazzola, {\em Phys. Rev.} {\bf187}, 1767 (1969).
\bibitem{Colpi} M. Colpi, S. L. Shapiro and I. Wasserman, {\em Phys. Rev. Lett.} {\bf57}, 2485 (1986).
\bibitem{Jetzer} P. Jetzer, {\em Phys. Rep.} {\bf220}, 163 (1992).
\bibitem{Jetzer1} P. Jetzer, P. Liljenberg and B.-S. Skagerstam, {\em Astropart. Phys.} {\bf1}, 429 (1993).
\bibitem{Henriques} A. B. Henriques and L. E. Mendes, {\em Astrophys. Space Sci.} {\bf300}, 367 (2005).
\bibitem{Henriques1} A. B. Henriques, R. Liddle and R. G. Moorhouse, {\em Nucl. Phys. B} {\bf337}, 737 (1990).
\bibitem{Hafizi}M. Hafizi, {\em Int. J. Mod. Phys. D} {\bf7}, 975 (1998).
\bibitem{Capozziello} S. Capozziello, G. Lambiase and D. F. Torres, {\em Class. Quantum Grav.} {\bf17}, 3171 (2000).
\bibitem{Ho} J. Ho, S. Kim and B.-H. Lee, {\em gr-qc/9902040}.
\bibitem{Alcubierre} M. Alcubierre, F. S. Guzm$\acute{a}$n, T. Matos, D. Nunez, L. A. Ure$\acute{n}$a-L$\ddot{o}$pez and P. Wiederhold, {\em Class. Quantum Grav.} {\bf19}, 5017 (2002).
\bibitem{Padmanabhan} T. Padmanabhan, {\em Phys. Reports} {\bf380}, 235 (2003).
\bibitem{AGReiss} A. G. Reiss {\em et al}, {\em Astron. J.} {\bf116}, 1009 (1998).
\bibitem{Schunck1}  F. E. Schunck, {\em astro-ph/9802258}.
\bibitem{Astefanesei} D. Astefanesei and E. Radu, {\em Nucl. Phys.} {\bf665}, 594 (2003).
\bibitem{Seidel02} E. Seidel and W. Suen, {\em Phys. Rev. D} {\bf42}, 384,(1990).
\bibitem{Seidel03} E. Seidel and W. Suen, {\em Phys. Rev. Lett.} {\bf72}, 2516 (1994).
\bibitem{VT01} P. C. Vaidya and R. Tikekar, {\em J. Astrophys. Astron.} {\bf3}, 325  (1982).
\bibitem{sk} S. Karmakar, S. Mukherjee, R. Sharma and S. D. Maharaj, {\em Pramana -J. Phys.} {\bf68}, 881 (2007).
\bibitem{SNB01} S. Mukherjee, B. C. Paul and N. K. Dadhich, {\em Class. Quantum Grav.} {\bf14}, 3475  (1997).
\bibitem{MSchunck} E. W. Mielke and F. E. Schunck, {\em Nucl. Phys. B} {\bf564}, 185 (2000).
\bibitem{Genzel} R. Genzel {\em et al}, {\em Astrophys. J.} {\bf472},153 (1996).
\bibitem{Torres} D. F. Torres, S. Capozziello and  G. Lambiase,  {\em Phys. Rev. D} {\bf62}, 104012 (2000).
\end{thebibliography}
\end{document}